%% file: atomic.tex
\documentclass[graybox]{svmult}

\usepackage{colortbl}
\usepackage{type1cm}        % activate if the above 3 fonts are
\usepackage{makeidx}         % allows index generation
\usepackage{graphicx}        % standard LaTeX graphics tool
\usepackage{multicol}        % used for the two-column index
\usepackage[bottom]{footmisc}% places footnotes at page bottom

\usepackage{newtxtext}       %
\usepackage[varvw]{newtxmath}       % selects Times Roman as basic font
\usepackage{microtype}

\makeatletter
\providecommand*{\toclevel@author}{0}%
\providecommand*{\toclevel@title}{0}%
\makeatother

\usepackage{hyperref}
\usepackage[numbers,sort&compress]{natbib}

\usepackage{tikz}
\usetikzlibrary{arrows,positioning}

\usepackage{booktabs}
\usepackage{subcaption}

\makeindex             % used for the subject index
\newcommand{\hh}[1]{\noindent{\emph{#1}}}
\newcommand{\abinitio}{\emph{ab initio}}
\newcommand{\cpp}{C\texttt{++}}
\begin{document}

\title*{Symbolic Regression in Materials Science: Discovering Interatomic Potentials from Data}
%\title*{Symbolic Regression in Materials Science}
\titlerunning{Discovering Interatomic Potentials with Symbolic Regression}
\author{Bogdan Burlacu, Michael Kommenda, Gabriel Kronberger, Stephan Winkler, Michael Affenzeller}
\authorrunning{Burlacu et al.}
\institute{
    Heuristic and Evolutionary Algorithms Laboratory\newline
    University of Applied Sciences Upper Austria\newline
    Softwarepark 11, 4232 Hagenberg, Austria
    %Bogdan Burlacu \at University of Applied Sciences Upper Austria, Address of Institute, \email{bogdan.burlacu@fh-ooe.at} \and
    %Michael Affenzeller \at University of Applied Sciences Upper Austria, Address of Institute, \email{michael.affenzeller@fh-ooe.at} \and
    %Michael Kommenda \at University of Applied Sciences Upper Austria, Address of Institute, \email{michael.kommenda@fh-ooe.at} \and
    %Stephan Winkler \at University of Applied Sciences Upper Austria, Address of Institute, \email{stephan.winkler@fh-ooe.at} \and
    %Gabriel Kronberger \at University of Applied Sciences Upper Austria, Address of Institute, \email{gabriel.kronberger@fh-ooe.at} \and
}
\maketitle

\abstract*{}

\abstract{
    Particle-based modeling of materials at atomic scale plays an important role in the development of new materials and understanding of their properties. The accuracy of particle simulations is determined by \emph{interatomic potentials}, which allow to calculate the potential energy of an atomic system as a function of atomic coordinates and potentially other properties. First-principles-based \abinitio{} potentials can reach arbitrary levels of accuracy, however their aplicability is limited by their high computational cost.
\newline\indent
Machine learning (ML) has recently emerged as an effective way to offset the high computational costs of \abinitio{} atomic potentials by replacing expensive models with highly efficient surrogates trained on electronic structure data. Among a plethora of current methods, symbolic regression (SR) is gaining traction as a powerful ``white-box'' approach for discovering functional forms of interatomic potentials.
%due to its natural representation facilitating the incorporation of physical concepts and its ability to produce interpretable models.
\newline\indent
This contribution discusses the role of symbolic regression in Materials Science (MS) and offers a comprehensive overview of current methodological challenges and state-of-the-art results. A genetic programming-based approach for modeling atomic potentials from raw data (consisting of snapshots of atomic positions and associated potential energy) is presented and empirically validated on \emph{ab initio} electronic structure data.
}
\newpage

\section{Introduction}\label{sec:introduction}

Materials Science (MS) is a highly interdisciplinary field incorporating elements of physics, chemistry, engineering and more recently, machine learning, in order to design and discover new materials.
The rapid increase in processing power over the last decades has made computational modeling and simulation the main tool for studying new materials and determining their properties and behavior.
Computational approaches can deliver accurate quantitative results without the need to set up and execute highly complex and costly physical experiments.

\begin{figure}
    \centering
    \includegraphics[width=0.8\textwidth]{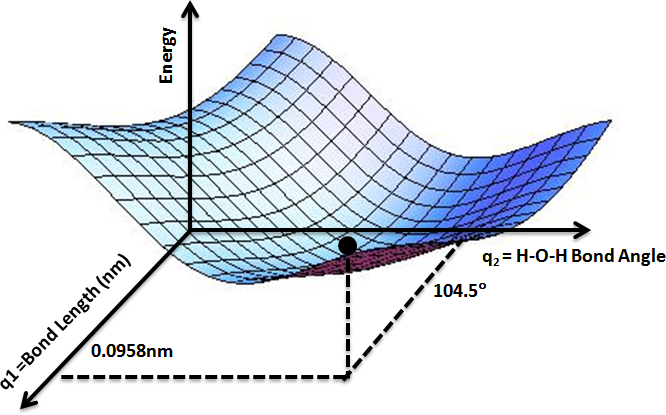}
    \caption{PES for water molecule: Shows the energy minimum corresponding to optimized molecular structure for water- O-H bond length of 0.0958nm and H-O-H bond angle of 104.5°. Image from Wikipedia \textcopyright{AimNature}}\label{fig:pes-example}
\end{figure}

Potential energy surfaces (PES), describing the relationship between an atomic system's potential energy and the geometry of its atoms, are a central concept in computational chemistry and play a pivotal role in particle simulations. An example PES for the water molecule is shown in Figure~\ref{fig:pes-example}. The mathematical function used to calculate the potential energy of a system of atoms with given positions in space and generate the PES is called an \emph{interatomic potential} function. The form it this function, it's physical fidelity as well as its complexity and efficiency are critical components in simulations used to predict material properties.

The ability to simulate large particle systems over long timescales depends critically on the accuracy and computational efficiency of the interatomic potential. Broadly speaking, the more accurate the methods, the lower its computational efficiency and the more limited its applicability.
For example, first-principle modeling methods such as \emph{density functional theory} (DFT)~\cite{Kohn1965} provide highly accurate results by considering quantum-chemical effects but are not efficient enough to simulate large systems containing thousands of atoms over long timescales of nanoseconds \cite{Rothe2019}.

Molecular dynamics (MD) simulations treat materials as systems consisting of many microscopic particles (atoms) which interact with each other through interatomic potentials depending mainly on their positions and are governed by the laws of statistical thermodynamics. Macroscopic properties of materials are obtained as time and/or ensemble averages of processes emerging at the microscopic scale~\cite{Hospital2015}. %The accuracy and efficiency of MD simulations depend entirely on the interatomic potentials used. %Despite this they are still popular in simulations under the name of \emph{(classical) force fields}.

Empirical and semi-empirical methods treat atomic interactions in a more coarse-grained manner via parameterized analytical functional forms and trade-off accuracy for execution speed in order to enable simulations at a larger scale. Although they are computationally undemanding, they are only able to provide a qualitatively reasonable description of chemical interactions~\cite{Unke2021}.

Machine learning (ML) interatomic potentials aim to bridge the gap between quantum and empirical methods in order to deliver the best of both worlds: functional forms that are as efficient as empirical potentials and as accurate as quantum-chemical approaches.

\subsection{Materials informatics and data-driven potentials}

Building upon the three established paradigms of science that have lead to many technological advances over time: experimental, theoretical and simulation-based, a fourth ``data-driven`` paradigm of science is emerging today using machine learning and the large amounts of experimental and simulation-data available~\cite{Agrawal2016}. ``Big-data'' science unifies the first three paradigms and opens up new avenues in materials science under the umbrella term of \emph{materials informatics}. The field of material informatics is very new and many unsolved questions still remain open and wait for proper answers~\cite{Himanen2019}.

Machine learning interaction models are generated on the basis of quantum-chemical reference data consisting of a series of snapshots of atomic coordinates, associated potential energy of the system and optionally other properties.

In molecular dynamics simulations, the system's potential energy is typically decomposed into a set of independent $m$-body interactions that are a function of each particle's position, $\mathbf{r}$. For a two-body or pair potential, it is assumed that the energy contributions from each pair of interacting particles are independent of other pairs and therefore:
\begin{equation}
    E = \sum_{\langle i, j \rangle} g(\mathbf{r}_i, \mathbf{r}_j)\label{eq:two-body-potential}
\end{equation}

For a three-body potential, triplets of atoms are also considered:
\begin{equation}
    E = \sum_{\langle i, j \rangle} g(\mathbf{r}_i, \mathbf{r}_j) + \sum_{\langle i, j, k \rangle} h(\mathbf{r}_i, \mathbf{r}_j, \mathbf{r}_k)\label{eq:three-body-potential}
\end{equation}

Traditionally, the functions $g$ and $h$ are represented by all kinds of empirical or semi-empirical analytical functions. With the advent of machine learning and data-based modeling, it becomes possible to automatically search for these functional forms with the help of ab initio training data. Substantial effort has already been put into this direction and many machine learning models have been successful in discovering interatomic potentials for a variety of chemical configurations~\cite{Pilania2021}.

\subsection{Current challenges}\label{subsec:current-challenges}

Despite their success in representing atomic interactions, ML-methods are not without their own challenges. Deriving highly-accurate and tractable analytic functional forms for high-dimensional PESs is a very active field of research.
%
%Currently available potentials are far from satisfying all the needs and deriving potentials that fulfill as many of these requirements is a very active field of research~\cite{Behler2017}:
The most important requirements for ML-based PESs are:
\begin{itemize}
    \item general applicability and absence of ad-hoc approximations (transferability)
    \item accuracy close to first-principles methods (including high-order many-body effects)
    \item very high efficiency to enable large simulations
    \item the ability to describe chemical reactions and arbitrary atomic configurations
    \item the ability to be automatically constructed and systematically improved
\end{itemize}

Currently available potentials are far from satisfying all the needs~\cite{Behler2016}, mainly due to the following difficulties and shortcomings:

\paragraph{\hh{Physical plausibility}}

Closed physical systems are governed by various conservation laws that describe invariant properties. These fundamental principles of nature provide strong constraints that can be used to guide the search towards physically-plausible ML models~\cite{Unke2021}. In molecular systems each conserved quantity is associated with a differentiable symmetry of the action of a physical system.

Typical conserved quantities include temporal and roto-translational invariance (i.e. total energy, linear and angular momentum).
Forces must be the negative gradient of the potential energy $E$ with respect to atomic positions $r_i$: $$F_i = -\nabla r_i E$$
When atoms move, they always acquire the same amount of kinetic energy as they lose in potential energy, and vice versa -- the total energy is conserved. The potential energy of a molecule only depends on the relative positions of atoms and does not change with rigit rotations or translations.

Another aspect of invariance is permutational invariance resulting from the fact that from the perspective of the electrons, atoms with the same nuclear charge appear identical to each other and can thus be exchanged without affecting the energy or the forces. To ensure physically meaninful predictions, ML-based models must exhibit the same invariant behavior as the true potential energy surface.

\paragraph{\hh{Accuracy}}

Accuracy is one of the most important requirements of ML potentials. The predicted energies and forces should be as close as possible to the underlying \emph{ab initio} data. Numerical accuracy of the ML models is restricted by the intrinsic limitations of their functional form and descriptors (input variables) used. For example, conceptual problems related to incorporating rotational, translational and permutational invariance into descriptors are of primary relevance~\cite{Ghiringhelli2015,Shao2016,Shapeev2016,Behler2016} as well as their optimal design~\cite{Gao2019}.

\paragraph{\hh{Transferability}}

Ideally, potentials should be generally applicable and should not be restricted to specific types of atomic configurations. Due to their mathematical unbiased form, ML methods are promising candidates to reach this goal. However in practice, developed potentials often perform very well in applications they have been designed for, but are too system-specific and thus cannot be easily transferred from one system to another. The issues of extensibility, generality and transferability of the ML potentials need to be explicitly addressed~\cite{Behler2016}.

\paragraph{\hh{Complexity and data requirements}}

Another issue worth mentioning here is the mathematical complexity of ML potentials. For example, the most popular ML methods used to represent many-body PESs, ANNs, require complex architectures with many adjustable parameters (weights of neural synapses and neuron biases) to yield sufficiently flexible and invariant PESs representations. For this, large amounts of training data (often dozens or even hundreds of thousands points) are needed.
%to avoid over-fitting issues. -- BB: not quite a good connection to overfitting
On the other hand, the number of training data should be kept as low as possible since they are are calculated via demanding quantum-chemical methods. It means that as simple as possible analytic representations of PESs are needed.

\paragraph{\hh{Integration of physical knowledge and interpretability}}

Related to the mathematical complexity issue, it is also important to note that most of the ML methods (e.g., ANN, SVM) are of a ``black-box'' nature and may be less amenable to including physical information into the functional forms, relying at least partially on physics-inspired features considered in atomic descriptors. This often leads to increased mathematical and computational complexity of resulting interaction models. One of the main directions of the current development in ML-based computational MS is the shift from ``black-box'' methods towards ``white-box''
%or at least``gray-box''
methods which often offer better interpretability.% of resulting formulas and are more amenable to the integration of physical knowledge into developed models.

\section{State of the art}\label{sec:state-of-the-art}

A plethora of machine learning approaches have recently emerged as a powerful alternative for finding a functional relation between an atomic configuration and corresponding energy~\cite{Handley2014,Behler2016,Dral2020}.

Several ML techniques such as polynomial fitting~\cite{Brown2004}, Gaussian processes~\cite{Bartok2013}, spectral neighbor analysis~\cite{Thompson2015}, modified Shepard interpolation~\cite{Ischtwan1994}, moment tensor potentials~\cite{Shapeev2016}, interpolating moving least squares~\cite{Kim2003}, support vector machines~\cite{Balabin2011}, random forests~\cite{Kim2016}, artificial neural networks (ANNs)~\cite{Shao2016,Zhang2018,Chen2020,Hey2020} or symbolic regression (SR)~\cite{Mueller2020} have been successfully employed for a variety of systems.

More detailed reviews of current ML potentials can be found for example in~\cite{Handley2014}, \cite{Mueller2014a}, \cite{Pilania2021} or \cite{Unke2021}. Particularly, ANNs have received a considerable attention and are probably the most popular form of ML potentials used in MS~\cite{Zhang2018}. However, methods based on symbolic regression are gaining in popularity due to the advantages they bring in solving aspects of physical knowledge integration, efficiency and interpretability~\cite{Makarov1998,Sastry2007,Slepoy2007,Brown2008,Brown2010,Belluci2011,Belluci2012,Mueller2014,Hernandez2019}.

In the following, we refer to symbolic regression in its canonical incarnation that employs genetic programming to perform a search over the space of mathematical expressions.
Symbolic regression approaches have succeeded in rediscovering simple forms of potentials that deliver qualitatively good results in a series of specific applications, some of which are described below.

\subsection{Directed search}\label{subsec:directed-search}

The goal of directed search is to improve search efficiency by limiting the hypothesis space to a functional form known to deliver qualitatively good results, instead of searching for a brand new potential.

Makarov and Metiu~\cite{Makarov1998} use the Morse potential as a functional template for modeling diatomic molecules (see Section~\ref{sec:appendix}, Eq.~\ref{eq:morse-potential}).
They rewrite it in the form:
\begin{equation}
    M\big( D(r), R(r) \big) = D(r) \big(1-\exp \big( R(r) \big) \big)^2\label{eq:mak-functional-template}
\end{equation}
and use genetic programming to find the best $D(r)$ and $R(r)$.

The directed search approach is augmented with an error metric that better reflects the physical characteristics of the problem.
A standard error metric such as the MSE has the disadvantage of overemphasizing high-energy points which are rarely used during simulation.
%\begin{equation}
%    F(a) = \sum_i \Big( E(r_i) - f(r_i; a) \Big)^2\label{eq:mak-error-metric}
%\end{equation}
For this reason, the authors found it advantageous to introduce a scaling factor:
\begin{equation}
    F(a) = \sum_i \frac{\big( E(r_i) - f(r_i; a) \big)^2 }{E^2(r_i) + \delta^2}\label{eq:mak-error}
\end{equation}
where the constant $\delta$ is added to prevent division by zero.

For each function $f_\alpha$ in the population of individuals, the fitness function is then defined as:
\begin{equation}
    p_\alpha = \exp \big( -\beta F_\alpha \big)\label{eq:mak-fitness}
\end{equation}
where parameter $\beta$ controls how discriminating the function is and is adaptively updated during the run. The search starts with a small value for $\beta$ which is gradually increased as the search improves.

The authors note the importance of including the derivative of the energy in the training data:
\begin{equation}
    F'(a) = \sum_i \frac{\left| \nabla E(r_i) - \nabla f(r_i; a) \right|^2}{\left| \nabla E(r_i) \right|^2 + \delta'^2}
\end{equation}
Leading to an expanded fitness function $p_\alpha$:
\begin{equation}
    p_\alpha = \exp \left( -\beta(F + F') \right)
\end{equation}

% The following is commented out because it's too detailed
%The following genetic operators are introduced, noting their tendency to be more disruptive than their genetic algorithm counterparts:
%\begin{itemize}
%    \item \emph{Reproduction} makes a copy of the current tree.
%    \item \emph{Mutation} replaces a tree with a new randomly-generated one.
%    \item \emph{Grafting} replaces a subtree with a randomly-generated one.
%    \item \emph{Scaling} replaces a variable $r$ with $r(1+R)$ where $R$ is a random number.
%    \item \emph{Parameter change} replaces a constant $a$ with $a+R$ where $R$ is a random number.
%\end{itemize}
%From the above, grafting, parameter change and scaling are applied with a higher probability, followed by mutation and crossover with a lower probability. As the fitness of the population improves, the relative frequency of parameter changes is increased.
%
The recombination pool is filled using a proportional selection scheme. An additional ``natural selection'' operator employs a ``badness list'' $b_\alpha = \exp(\beta F_\alpha)$ whose elements are the inverse of the fitness. Old individuals are replaced with a probability proportional with badness.
%
%Optimization of function parameters is not performed indiscriminately but is instead applied with a probability proportional to fitness, using a steepest descent algorithm.
%
\paragraph{\emph{Results}}

The directed search approach is is shown to perform better than an undirected search over the search space, on training data generated using the Lippincott potential (Section~\ref{sec:appendix}, Eq.~\ref{eq:lippincott-potential}).
A population size of 500 individuals is evolved over 150 generations (75,000 evaluations) using the primitive set $\mathcal{P} = \{ +, -, \div, \times, \exp \}$.
Furthermore, a search directed by a Lennard-Jones potential gives accuracy comparable to that directed by a Morse function, suggesting that restricting the hypothesis space with an appropriate functional template is a powerful and general approach in the search for interatomic potentials. In the case of the Lennard-Jones potential (Section~\ref{sec:appendix}, Equation~\ref{eq:lennard-jones-potential}) the functional template was defined as
\begin{equation}
f(r) = 4 D(r) \Bigg[ \frac{1}{4} + \Big(\frac{1}{R(r)}\Big)^{12} - \Big(\frac{1}{R(r)}\Big)^6 \Bigg]
\end{equation}

The authors additionally note that some of the returned models, although accurate, exhibited unphysical behavior and did not extrapolate well. For example, one of the returned models based on the Lennard-Jones functional form had very good accuracy but contained a singularity at $r=12$ \AA, a point outside the interpolation range. The authors address overfitting by fitting the parameters of both the energy function and its derivative in the local search phase. This reduces the chance of obtaining pathological curves in the model extrapolation response.

Finally, Makarov and Metiu also model the potential of a triatomic molecule on \emph{ab initio} data consisting of 60 nuclear configurations, showing that directed search maintains high levels of accuracy and scales favorably with dimensionality.

% ------------------------------------------------------------------------------------- %

\subsection{Directed Search with Parallel Multilevel Genetic Program}\label{subsec:directed-search-pmgp}

Belluci and Coker~\cite{Belluci2011,Belluci2012} employ symbolic regression to discover empirical valence bond (EVB) models using directed search augmented with a multilevel genetic programming approach: the lower level (LLGP) optimizes co-evolving populations of models, while the higher level (HLGP) optimizes genetic operator probabilities of the lower level populations. The approach entitled Parallel Multilevel Genetic Program (PMLGP) found accurate EVB models for proton transfer in 3-hydroxy-gramma-pyrone (3-HGP) in gas phase and protic solvent as well as ultrafast enolketo isomerisation in the lowest singlet excited state of 3-hydroxyflavone (3-HF).

At the lower level (LLGP), the authors use the same error metric and fitness as in~\cite{Makarov1998}, namely Equations \ref{eq:mak-error} and \ref{eq:mak-fitness}. LLGP individuals represent the $R(r)$ functional part of the Morse potential (see Equation~\ref{eq:mak-functional-template}).
Remarkably, PMLGP does not use crossover but instead uses six different mutation operators:
\begin{itemize}
    \item \emph{Point mutation} randomly replaces a subtree with a randomly-generated one.
    \item \emph{Branch mutation} replaces a binary operator with one of its arguments at random.
    \item \emph{Leaf mutation} replaces a leaf node with another randomly selected leaf.
    \item \emph{New tree mutation} replaces an entire tree with a newly generated tree.
    \item \emph{Parameter change} replaces each parameter value $a_i$ with $a_i + (R-0.5)\gamma$, where $R$ is a uniform random number on the unit interval and $\gamma$ is a scaling constant.
    \item \emph{Parameter scaling} replaces each parameter value $a_i$ with $a_iR\gamma$, where $R$ is a uniform random number on the unit interval and $\gamma$ is a scaling constant.
\end{itemize}
Of the last two types of mutation, parameter change is designed to make small local moves in parameter space, while parameter scaling is designed to make large moves in parameter space as to escape the basins of attraction of local optima.
Selection is performed using stochastic universal sampling~\cite{Baker1987}.

At the higher level (HLGP) a real vector encoding is used to represent genetic operator probabilities. The population is initialized with $k$ random vectors $P_k = \left(p_1^{(k)}, ..., p_6^{(k)} \right)$, with $\sum_i p_i^{(k)} = 1$, where $k$ ranges from $1$ to the total number $N_p$ of processors, such that each vector corresponds to one of the LLGP populations whose operator probabilities it dynamically adapts.

The fitness of each vector $P_k$ is evaluated based on the maximum fitness delta in the corresponding LLGP population over a specified time interval $\Delta t$:
\begin{equation}
    F_k^\text{HLGP} = \frac{\Delta F_{\max}^\text{LLGP}}{\Delta t}
\end{equation}
This is based on the idea that the larger the magnitude of $F_k^\text{HLGP}$, the more successful the set of probabilities $P_k$ at improving the fitness of the population.

Two genetic operators are used to modify the probability vectors $P_k$:
\begin{itemize}
    \item \hh{Mutation} changes each component of the vector by a random amount with the constraint that all components sum up to one. This operator kicks in when the fitness of a vector $P_k$ drops below a given threshold.
    \item \hh{Adaptation} attempts to improve the probability distribution given by $P_k$ by using feedback from the LLGP. Each LLGP builds a histogram of the number of times each mutation produced the most fit member of the population. Then the success frequency of the mutation operator is given by:
        \begin{equation*}
            s_i = \frac{w_i m_i}{n},\ w_i = \frac{1}{p_i},\ n = \sum_i m_i
        \end{equation*}
        Here, $w_i$ is a weight, $m_i$ is the number of successful mutations for the $i^\text{th}$ operator (component of $P_k$) and $n$ is the total number of successful mutations (for all operators).
        Based on the success frequencies, adaptation shifts a random amount of probability from the least successful operator to the most successful operator.
\end{itemize}

The number of LLGP populations (and HLGP individuals, respectively) is set to the number of available processors. Initially, all LLGP populations are identical but diverge during evolution as each corresponding fitness function is parameterized with a different value of $\beta$ evenly sampled over a specified range.
In effect, this applies different selection pressures on each LLGP population.
Migrations are performed after the last adaptation step in HLGP. At this point, copies of the fittest individual in each LLGP population are sent to all the other populations, where they replace the least fit individual.

\paragraph{\emph{Results}}

Training data for five different diatomic molecules (CO, H$_2$, HCl, N$_2$, O$_2$) was generated using differently parameterized Morse functions, Gaussian functions and double well functions.
The corresponding directed search spaces are given by:
\begin{align}
    F_M &= D \big( 1 - \exp(-R(r;a)) \big) ^ 2 + c & \text{Morse}\\
    F_G &= A \exp \big( R(r;a)^2 \big) & \text{Gaussian}\\
    F_D &= D_1 \big( 1 - \exp(-R_1(r;a)) \big) ^ 2 + D_2 \big( 1 - \exp(-R_2(r;a)) \big) ^ 2 & \text{Double well}
\end{align}
Parameters $D, c, A, D_1$ and $D_2$ are optimized by including them as leaves in the trees.

The PMLGP approach was compared against a standard parallel genetic programming implementation (SPGP). In both cases, populations of 500 individuals were evolved in parallel on 8 processors for 20,000 generations.
The function set $\mathcal{F} = \{ +, -, \times, \div, \exp \}$ was used for internal nodes and the terminal set $\mathcal{T} = \{ r, a_1, ..., a_{10} \}$ was used for the leaf nodes.

PMLGP was shown to converge faster and achieve higher accuracy than SPGP. The obtained model of the EVB surface accurately reproduced global features of the \emph{ab initio} data. The approach provides a basis for high quality many-body potentials for studying gas and solution phase photon reactions.

% ------------------------------------------------------------------------------------- %

\subsection{Parallel tempering}\label{subsec:parallel-tempering}

Slepoy, Peters and Thompson~\cite{Slepoy2007} use a hybrid approach consisting of genetic programming, Monte Carlo sampling and parallel tempering to discover the functional form of the Lennard-Jones pair potential.

Parallel tempering is an approach for parallel genetic programming where several islands (or \emph{replicas}) evolve at a different effective temperature. High effective temperatures favor exploration by accepting new trees even if their fitness is poor, low effective temperatures favor exploitation by being sensitive to small changes in fitness. By using replicas at different temperatures the approach simultaneously performs both exploitation and exploration.

The remarkable aspect about this approach is that it marks the first large-scale application of genetic programming in materials science with interesting extensions to the canonical Koza-style algorithm and without restrictions of the hypothesis space.

The training data used consists of 10 nuclear configurations of 10 particles placed in 3-d space. The Lennard-Jones potential describes the interations between pairs of particles, therefore a nuclear configuration's energy is given by the sum of pairwise potentials:
\begin{equation}
    E_\text{conf} = \sum_{<i,j>} V_\text{LJ}(r_{ij})
\end{equation}
where $r_{ij} = \|\mathbf{r}_i, \mathbf{r}_j\|$ is the distance between particles $i$ and $j$. Fitness is defined as the negative mean squared error.
%
%The negative mean squared error is used as a fitness measure:
%\begin{equation}
%    F = -\frac{1}{N} \sum_{i=1}^{N} \big(\hat{E}_i - E_i \big) ^2
%\end{equation}

The evolutionary search is organized as a three-stage process consisting of: generation, mutation and testing.
%\begin{itemize}
%    \item \emph{Generation}. This stage creates the new population by pass-through or crossover, with equal probability. Pass-through simply copies the fittest tree from the old population in the new one, with the restriction that a tree can only be passed through once. Crossover generates a new tree from two parents with the restriction that crossover points in each tree cannot be at the root or on the last depth level. The parents are selected using tournaments of size four.
%
%    \item \emph{Mutation}. Each new tree is either left unchanged or mutated with equal probability. The subtree to be mutated is uniformly chosen from all available nodes and replaced with a new randomly-generated subtree. The uniform selection is an intentional design choice meant to introduce a bias towards leaf nodes and therefore small changes to parameters instead of big functional changes.
%
%    \item \emph{Testing}. During the testing phase new individuals are compared against the old ones by simultaneous iteration over the two lists. The old trees list is ordered by fitness while the new trees list is filled in creation order.
%\end{itemize}

Offspring individuals are tested for acceptance into the new population. A new tree is unconditionally accepted if its fitness exceeds the old one at the same index. Otherwise, it is accepted with the Boltzmann probability:
        $$
        P_\text{accept} = \min \Bigg\{ 1, \exp\Bigg( \frac{F_\text{new} - F_\text{old}}{T} \Bigg) \Bigg\}
        $$
        where $F_\text{old}$ and $F_\text{new}$ are the old and new fitness values, and $T$ is the effective temperature.

After each generation, each sub-population exchanges one tree with its left neighbour in temperature space and one tree with its right neighbor. The trees to be swapped are selected with equal probability from their respective populations. The tree swap is accepted with a probability based on the relative Boltzmann weights of the two trees.
$$
    P_\text{acc} = \min \Bigg\{ 1, \exp \left[ \left( \frac{1}{T_i} - \frac{1}{T_{i+1}} \right) \Big( F_{i+1} - F_{i} \Big) \right] \Bigg\}
$$

\paragraph{\emph{Results}}

A large scale experiment was performed on a cluster made of 100 AMD Opteron 2.2 Ghz processors. The trees were restricted to minimum depth 3 and maximum depth 4. 200 replicas with temperatures distributed logarithmically from 0.1 to 10 were used. The replica size was chosen to be either $N=\text{10,000}$ or $N = \text{50,000}$ individuals. The primitive set consists of elementary operations $\mathcal{P} = \{ +, -, \times, \div, \exp, \mid \cdot \mid \}$.

The proposed approach successfully discovered the Lennard-Jones potential or arithmetic equivalents within 100 generations. Interestingly, the expended effort was estimated to be somewhere in the range of $10^9$ evaluated trees, which represents only a small fraction of the possible trees with depth 4 (around $2.9 \times 10^{36}$)~\cite{Slepoy2007}.

A number of ideas for improving the physical fidelity of the developed functional forms and their generality and transferability are suggested
\begin{itemize}
    \item Inclusion of additional properties and forces on individual atoms in the training set
    \item Primitive set extension to include three-body interactions
    \item Integration of physical knowledge (inclusion of symmetries, invariances)
\end{itemize}

% ------------------------------------------------------------------------------------- %

\subsection{Symbolically-Regressed Table KMC}\label{subsec:srt-kmc}

In order to increase the time scale of simulations, molecular dynamics can be combined with kinetic (dynamic) Monte Carlo (KMC) techniques \cite{Binder1993} that coarse-grain the state space, for example via discretization (e.g. assign an atom to a lattice site). The main assumption is that multiscale modeling requires only relevant information at the appropriate length or time scale.

KMC constructs a look-up table consisting of an \emph{a priori} list of events such as atomic jumps or off-lattice jumps. This yields several order of magnitude increases in simulated time and allows to directly model many processes unaproachable by MD alone. However, identifying barrier energies from a list of events is difficult and restricts the applicability of the method.

Here, symbolic regression is proposed to identify the functional form of the potential energy surface at barrier energy points from a limited set of \emph{ab initio} training data.% (via semi-empirical, tight-binding or \emph{ab initio} potentials).
The method entitled Symbolically-Regressed Table KMC (sr-KMC) \cite{Sastry2007} provides a machine learning replacement for the look-up table in KMC, thus removing the need of explicit calculation of all activation barriers.
%previous limitations on the number of barrier points and increasing the number of active configurations that can be simulated.

Sastry~\cite{Sastry2007} show that symbolic regression allows atomic-scale information (diffusion barriers on the potential energy surface) to be included in a long-time kinetic simulation without maintaining a detailed description of the all atomistic physics, as done within molecular dynamics.

In this approach, ﬁtness is computed as a weighted mean absolute error between the predicted and calculated barriers, for $N$ random configurations:

\begin{equation}
F = \frac{1}{N} \sum_{i=1}^N w_i \left| \Delta E_\text{pred}(\vec{x}_i) - \Delta E_\text{calc}(\vec{x}_i)  \right|
\end{equation}

Setting $w_i = | \Delta E_\text{calc} |^-1$ gives preference to predicting accurately lower-energy (most significant) events over higher energy events.

The algorithm uses the \emph{ramped-half-and-half} tree creation method, tournament selection and Koza-style subtree crossover, subtree mutation and point mutation~\cite{Koza1992}.

\paragraph{\emph{Results}}

sr-KMC is applied to the problem of vacancy-assisted migration on the surface of phase-separating $Cu_xCo_{1-x}$ at a concentrated allow composition ($x = 0.5$). Two types of potentials (Morse and TB-SMA) are used to generate the training data via molecular dynamics.
%The system consists of five layers with 100 to 625 atoms in each later. First and second nearest-neighbor (n.n.) jumps are considered.
%
The number of active configurations is limited knowing that only atoms in the environment locally around vacancy and migrating atoms significantly influence the barrier energies.

The inline barrier function is represented from the primitive set $\mathcal{P} = \mathcal{F} \cup \mathcal{T}$, with $\mathcal{F} = \{ +, -, \times, \div, \text{pow}, \exp, \sin \}$ and $\mathcal{T} = \{ \vec{x}, \mathcal{R} \}$. Here, $\vec{x}$ represents the current active configuration and $\mathcal{R}$ is an emphemeral random constant.

The results show that GP predicts all barriers within $0.1\%$ error while using less than $3\%$ of the active configurations for training. This leads to a significant scale-up in real simulation time and a significant reduction in the CPU time needed for KMC.
sr-KMC is also compared against the basic KMC approach (using a table look-up) where it was shown to perform orders of magnitude faster.

The authors note that standard basis-set regression methods are generally not competitive to GP due to the inherent difficulty in choosing appropriate basis functions and show that quadratic and cubic polynomials perform worse in terms of accuracy (within $2.5\%$ error) while requiring energies for $\sim 6\%$ of the active configurations.

They also note that GP is robust to changes in the configuration set, the order in which configurations are used or the labeling scheme used to convert the configuration into a vector of inputs.

% ------------------------------------------------------------------------------------- %

\subsection{Hierarchical Fair Competition}\label{subsec:hierarchical-fair-competition}

Brown, Thompson and Schultz~\cite{Brown2008,Brown2010} are able to rediscover the functional forms of known two- and three-body interatomic potentials using a parallel approach to genetic programming with extensions towards better generalization. Their implementation is based on  Hierarchical Fair Competition (HFC) by Jianjun et al.~\cite{Jianjun2005}.

The HFC framework~\cite{Jianjun2005} is designed towards maintaining a continuous supply of fresh genetic diversity in the population and protecting intermediate individuals who have not reached their evolutionary potential from being driven to extinction by unfair competition. It implements these goals with the help of a hierarchical population structure where individuals only compete with other individuals of similar fitness.

Brown et al. note that a correlation-based fitness measure would increase the efficiency of the search and propose the following formula using the Pearson correlation coefficient:
\begin{equation}
    F = \frac{N}{N + 100 - 100 \left| \displaystyle \sum_{i=1}^{N} \dfrac{(y_i - \bar{y})(\hat{y}_i - \bar{\hat{y}})}{p_i \sigma_y \cdot p_i \sigma_{\hat{y}}} \right|}
\end{equation}
Here, $N$ is the number of configurations and $p_i$ is the number of terms in the summation over $g$ (see Equation~\ref{eq:two-body-potential}).
Ordinary least squares is then used to fit the prediction $\hat{y}$ to the data by introducing scale and intercept terms to the functions $g$ and $h$:
\begin{equation}
    E = \sum_{\langle i, j \rangle} \left( a \cdot g(\mathbf{r}_i, \mathbf{r}_j) + b \right) + \sum_{\langle i,j,k \rangle} \left( c \cdot h(\mathbf{r}_i, \mathbf{r}_j, \mathbf{r}_k) + d \right)
\end{equation}
%The values of $a,b,c,d$ can be calculated analytically using least squares regression.

The approach is implemented in \textsc{pm-dreamer}, an open-source software package developed on top of the Open Beagle library for evolutionary computation~\cite{Gagne2006}, using its available genetic operators. These include several mutations (standard, shrink, swap, constant), subtree-swapping crossover, tournament selection and elitism:
\begin{itemize}
    \item \emph{Standard} mutation replaces a node in the tree with a randomly-generated subtree.
    \item \emph{Swap} mutation swaps two nodes in the tree.
    \item \emph{Shrink} mutation replaces a subtree with one of its arguments.
    \item \emph{Swap subtree} mutation swaps a subtree's arguments.
    \item \emph{Ephemeral} mutation changes the value of a constant in the tree.
\end{itemize}

Additionally, \textsc{pm-dreamer} implements support for distributed evolution using the MPI standard and introduces migration operators that exchange individuals between sub-populations at fixed intervals.

Bloat reduction strategies are implemented to prevent the expression trees from becoming increasingly large, a tendency observed especially in the case of three-body modeling. Two strategies are tested:
\begin{itemize}
    \item Using a simplification operator which replaces subtrees that evaluate to a constant value with the constant value. This operator is applied generationally at a fixed interval.
    \item Using penalty terms to the fitness function: in this case the fitness is decreased based on a threshold penalty size value $s_b$ and a maximum penalty size $s_e$, such that trees with length $< s_b$ are not penalized at all, and trees with length $> s_e$ are penalized fully (fitness is set to zero).
\end{itemize}

\hh{Local search}. %In order to improve the fit of the tree individuals, local search is employed with a set probability on the expression trees. Either a single constant or all constants in the expression are optimized. The local optimization employed is the derivative-free Nelder-Mead simplex algorithm.
Local search based on the derivative-free Nelder-Mead simplex algorithm is employed with a set probability, optimizing either a single constant or all the constants in the expression.
\smallskip

\hh{HFC Extension}. Brown et al. implement HFC in a parallel manner by allowing populations with different fitness thresholds to evolve in parallel, with periodic migrations between them. After migrations, populations that grow too large are ``decimated'' by removal of the least fit individuals, while populations that grow too small are supplemented with new randomly-generated individuals.

The population fitness thresholds are adapted during the search using two strategies: the first strategy uses a percentile parameter $p$ which determines the fitness threshold such that $p$ percent of individuals have equal or lower fitness. The second strategy uses fixed thresholds determined by the first non-zero threshold along with a scaling parameter equal to the ratio between successive thresholds.

\paragraph{\emph{Results}}

Training data for two- and three-body interactions was generated using the Lennard-Jones and Stillinger-Weber potentials (see Section~\ref{sec:appendix} Eqs.~\ref{eq:lennard-jones-potential}, \ref{eq:stillinger-weber-potential}).
In both cases, five configurations were used for training and 50 configurations were used for testing, in order to realistically represent the problem of obtaining models for condensed phases from a small training set.
%with a small training set and evaluate model overfitting and transferability.
The generated data includes pairwise distances between atoms, the energy and the force on a single atom.

%For the two-body interactions, experimental data was generated using the Lennard-Jones potential with $\sigma$ and $\epsilon$ set to 1. A cutoff of 2.5 was used such that no interaction with interatomic distance greater than 2.5 contributed to the energy.
%
%The training data consisted of five configurations with 55-65 pair interactions each, while the test set consisted of 50 random configurations with 52-104 pair interactions each. The discrepance between training and test set sizes was intentional, in order to realistically represent the problem of obtaining models for condensed phases with a small training set and evaluate model overfitting and transferability.
%
%For the three-body interactions, the Stillinger-Weber potential was used (see Section~\ref{sec:appendix}, Eq.~\ref{eq:stillinger-weber-potential}) in both its normal form and a simplified form where the smooth cutoff functions have been removed.
%
%The parameter values for the potential were taken from the original Stillinger-Weber potential for silicon. The training data consisted of five configurations and the test data consisted of 50 configurations. Each configuration was obtained as a random snapshot of a molecular dynamics simulation of 64-atom liquid Si. In each configuration, the number of two-body interactions ranged from 218 to 253 and the number of three-body interactions ranged from 1309 to 1794.

The authors compare a standard parallel island-based evolutionary model  against a parallel HFC evolutionary model, using 32 islands with a population of 10 000 individuals each, evolved over a period of 100 generations.

The primitive set used was $\mathcal{P} = \{ +, -, \times, \div, \text{pow}, \exp, \log, |\cdot| \}$, tournament selection was used with a tournament size of 6 and, in the case of the standard evolutionary model, 500 individuals were migrated between islands every 5 generations. For the HFC evolutionary model, the migration took place every generation, the first fitness threshold was set to 0.1 and the threshold ratio was set to 1.0. A detailed description of the other algorithm parameters is given in~\cite{Brown2010}.

After an initial tuning phase, the authors note that the number of interactions per energy point greatly increases the runtime requirements for optimization. The \cpp{} implementation of \textsc{pm-dreamer} is capable of doing vectorized evaluation of two- and three-body interatomic potential models using SIMD instructions in a manner similar to batched tree interpretation more typically used in GP. With vectorization, the evolutionary algorithm was found to perform roughly four times faster.

Local search was performed with varying probability on all constants in an expression using a maximum of 6 iterations of the Nelder-Mead algorithm. Simplification is performed every 20 generations.

Overall, the authors show that the HFC strategy consistently outperforms the standard generational evolutionary strategy and is able to find very accurate approximations for the targeted empirical potentials (Lennard-Jones and Stillinger-Weber).

% ------------------------------------------------------------------------------------- %

\subsection{Potential Optimization by Evolutionary Techniques (POET)}\label{subsec:poet}

POET~\cite{Hernandez2019} distinguishes itself from previously described approaches through an extended primitive set which includes summation symbols that aggregate local energy values around each atom, smoothing functions meant to exploit the ``short-sightedness'' of atomic interactions as well as leaf nodes representing the atomic neighborhood interaction radius.

The primitive set used by the algorithm consists of the function set $\mathcal{F} = \{ \sum, f, +, -, \times, \div, \text{pow} \}$ and terminal set $\mathcal{T} = \{ \mathcal{R}, r \}$. Here, $\sum$ are summation symbols, $f$ are smoothing symbols, $\mathcal{R}$ represents an ephemeral constant and, like before, $r$ represents the distance between atoms. Distances are considered within the neighborhood of each atom according to inner and outer cutoff radii $r_\text{in}$ and $r_\text{out}$.

\begin{figure}
    \centering
    \input{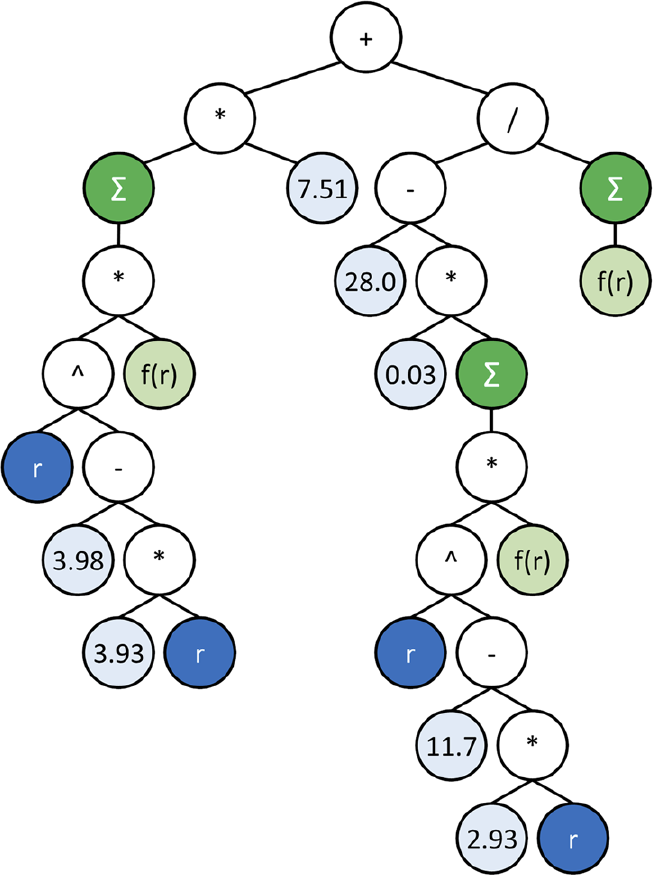}
    \caption{Interatomic potential obtained by Hernandez et al.~\cite{Hernandez2019}, representing local energy $E_i$ around the $i$th atom in electron volts. The expression resembles that of an embedded atom model with a pairwise repulsive term and a many-body attractive term formed by a non-linear transformation of pairwise interactions.}\label{fig:poet-example}
\end{figure}

An exemplary POET-tree including the special symbols $\sum, f$ and $r$ is shown in Figure~\ref{fig:poet-example}. This tree corresponds to the following function which returns the predicted value of the local energy $E_i$ around the $i$th atom considering the distances $r_{ij}$ to its neighbors:
\begin{align*}
    E_i &= 7.51 \sum_j r_{ij}^{3.98 - 3.93 r_{ij}} f(r_{ij})\\
        &+ \left( 28.01 - 0.03 \sum_j r_{ij}^{11.73-2.93 r_{ij}} f(r_{ij}) \right) \left( \sum_j f(r_{ij}) \right)^{-1}
\end{align*}

Hernandez et al. employ a parallel version of genetic programming where twelve populations are evolved simultaneously. The recombination pool in each population is filled from three separate sets of models: a set from the current population, a global set maintained with the overall best (non-dominated) individuals with regard to fitness, complexity and execution speed, and a set of individuals from the other populations. These sets are periodically filled up with individuals at preset intervals.

New individuals are generated by means of crossover and mutation. Crossover replaces a random subtree in the root parent with either another random subtree from another parent or with a linear combination of random subtrees from two different parents. The first method was applied with probability 0.9 while the second method was applied with probability 0.1.

Mutation can replace a subtree with a randomly generated one, swap the arguments of non-commutative symbols or change the symbols of function nodes.
Tree initialization is done using Koza's ramped-half-and-half method where the tree depth is sampled from a Gaussian distribution with $\mu = 5$ and $\sigma = 1$.

Local optimization of model coefficients is performed online during the run with the help of a covariance matrix adaptation evolution strategy (CMA-ES) optimizer and a conjugate gradient (CG) optimizer. CMA-ES is used to optimize the coefficients of models in the global set every 10,000 crossover and mutation operations. The CG algorithm is used to perform one optimization step for every newly generated individual.

\paragraph{\hh{Results}}

The proposed approach is validated using training data from DFT molecular dynamics simulations containing snapshots of atomic positions, energies, forces and stresses for an atomic system of 32 Cu atoms. The fitness measure is an aggregation of the energy, force and stress errors:
\begin{equation*}
    F = 1000 \cdot \left( 0.5 \text{MSE}_\text{energy} + 0.4 \text{MSE}_\text{force} + 0.1 \text{MSE}_\text{stress} \right)
\end{equation*}

The authors demonstrate POET's ability to rediscover Lennard-Jones and Sutton-Chen potentials. The generated models displayed low overfitting and high generalization being able to maintain high predictive accuracy for properties on which they were not trained. The simplicity of the models allows them to predict energies with speeds in the order of microseconds per atom, about 1-4 orders of magnitude faster than other ML potentials.
The authors also note that such simple models bring the additional advantage of requiring relatively small amounts of training data.

In terms of runtime performance of POET itself, the authors report 330 CPU-hours spent finding the exact Lennard-Jones potential, 3600 CPU-hours finding the exact Sutton-Chen potential and 360 CPU-hours to find the three best performing models reported in~\cite{Hernandez2019}. POET code is open-source and available online\footnote{\url{https://gitlab.com/muellergroup/poet}}.

%The method maintains a global set of individuals which are selected based on fitness, complexity and execution speed.

\subsection{Other applications}\label{subsec:other-applications}

Makarov and Metiu~\cite{Makarov2000} use directed genetic programming to find analytic solutions to the time-independent Schrödinger equation. The training data is generated by inverting the Schrödinger equation such that the potential is a functional depending on the wave function and the energy.

Kenoufi and Kholmurodov~\cite{Kenoufi2005} used symbolic regression to rediscover the Lennard-Jones potential and discovered a new potential for an argon dimer, using \abinitio{} data from DFT simulations.

Mueller et al.~\cite{Mueller2014} used symbolic regression for discovering relevant descriptors in hydrogenated nanocrystalline silicon with very low crystalline volume fraction, with applications in improving optical absorption efficiency in thin-film photovoltaics.

Wang et al.~\cite{Wang2019} used symbolic regression to discover the Johnson-Mehl-Avrami-Kolmogorov transformation kinetics law in the recrystallization process of copper; and the Landau free energy functional form for the displacive tilt transition in perovskite LaNiO$_3$.

Eldridge et.al~\cite{Eldridge2022} used the NSGA-III algorithm to learn interatomic potentials for carbon. The approach considered training error, individual age and individual complexity as objectives and was able to find simple and accurate potential functions.

% Wei et. al~\cite{Wei2020} used genetic algorithms to discover ``unexpected'' enhancement of thermal conductivity by disorder in 2D nanoporous graphene. The approach optimizes the degree of disorder of pores using thermal conductivity as an optimization target.

\subsection{Summary discussion}\label{subsec:summary-discussion}

State-of-the-art SR approaches for modeling interatomic potentials recognize the need for domain-specific extensions and hybridizations towards promoting physical plausibility and achieving high accuracy, while making the most out of the usually scarce quantities of available \emph{ab initio} training data.

Several extensions and hybridizations are used to augment the classic (Koza-style) genetic programming algorithm and increase its search performance. Parallel, island-based approaches are employed in all of the discussed methods, on the one hand, to more efficiently search the hypothesis space and on the other hand, to achieve higher throughput and alleviate the high computational costs of summations over two- and three-body atomic interactions.

Physical plausibility is promoted by restricting the hypothesis space (directed search), including domain specific information into the fitness function (e.g. weighing down high-energy points) or including additional targets (forces and stresses).

The results achieved so far have demonstrated the ability of symbolic regression to discover highly accurate and physically-plausible functional forms which can increase, due to their simplicity and efficiency, the performance of particle simulations (allowing them to run at larger scales or for longer times). At the same time, since the models are inherently more simple than similar black-box models such as ANNs, they tend to require a lesser amount of training data, which increases their applicability.

Overall, it can be concluded that symbolic regression represents a very promising approach for discovering more accurate and efficient potentials. However, designing evolutionary systems for this application area requires consideration of specific challenges as described in Section~\ref{subsec:current-challenges}.
In the following section we discuss several ideas towards a GP system design which is able to address the domain-specific requirements of interatomic potential.

\section{Designing GP for modeling interatomic potentials}

The problem of modeling interatomic potentials from data has the main particularity that data comes in the form of atomic configuration snapshots. Each configuration describes the positions of the atoms, its energy and optionally other properties (forces, stresses). Canonically, these data snapshots are generated by molecular dynamics simulation packages such as LAMMPS~\cite{Plimpton1995} or VASP~\cite{Kresse1996} and come in specific formats, see e.g. POSCAR\footnote{\url{https://www.vasp.at/wiki/index.php/POSCAR}}.

At the minimum, raw data has the following form:
$$
\begin{bmatrix}
    \mathbf{r}_1^{(1)} & \cdots & \mathbf{r}_M^{(1)} & E^{(1)}\\
    \vdots & & \vdots & \vdots\\
    \mathbf{r}_1^{(N)} & \cdots & \mathbf{r}_M^{(N)} & E^{(N)}
\end{bmatrix}
$$ where $\mathbf{r}_i^{(k)}$ is the position of the $i$th atom in the snapshot $k$ and $E^{(k)}$ is the associated potential energy value.

Since atomic interactions are computed based on the distances between atoms, the Cartesian coordinates need to be converted into sets of pairwise distances relative to each atom. It is then the role of the genetic system to evolve an accurate functional relationship between distances $r_{ij}$ and potential energy. For each training sample $k$, the symbolic regression model needs to process a set of pairwise atomic distances into a prediction for the energy with the help of summation symbol $\sum$.

As it becomes apparent from studying previous approaches described in Section~\ref{sec:state-of-the-art}, modeling interatomic potentials is a non-trivial problem which requires substantial computational resources. Previous implementations employed different strategies for parallelism as well as other optimization techniques such as vectorized model evaluation in order to speed up the search. Additionally, most approaches employed local search in order to improve model coefficients during evolution.

For this reason, we opt to extend the framework Operon~\cite{Burlacu2020} with additional functionality for modeling interatomic potentials. Operon already benefits from a fine-grained parallelism model designed for scalability and was shown to perform well on a variety of symbolic regression problems~\cite{LaCava2021}. Additionally, it features support for local optimization using the Levenberg-Marquardt optimization algorithm, where the gradient is obtained via automatic differentiation.

We adopt a multi-objective approach based on the NSGA2 algorithm~\cite{Deb2002} where model length is used alongside prediction accuracy in order to promote parsimony, interpretability and generalization.

\subsection{Symbolic regression in Operon}

\begin{figure}
    \includegraphics[width=1.0\textwidth]{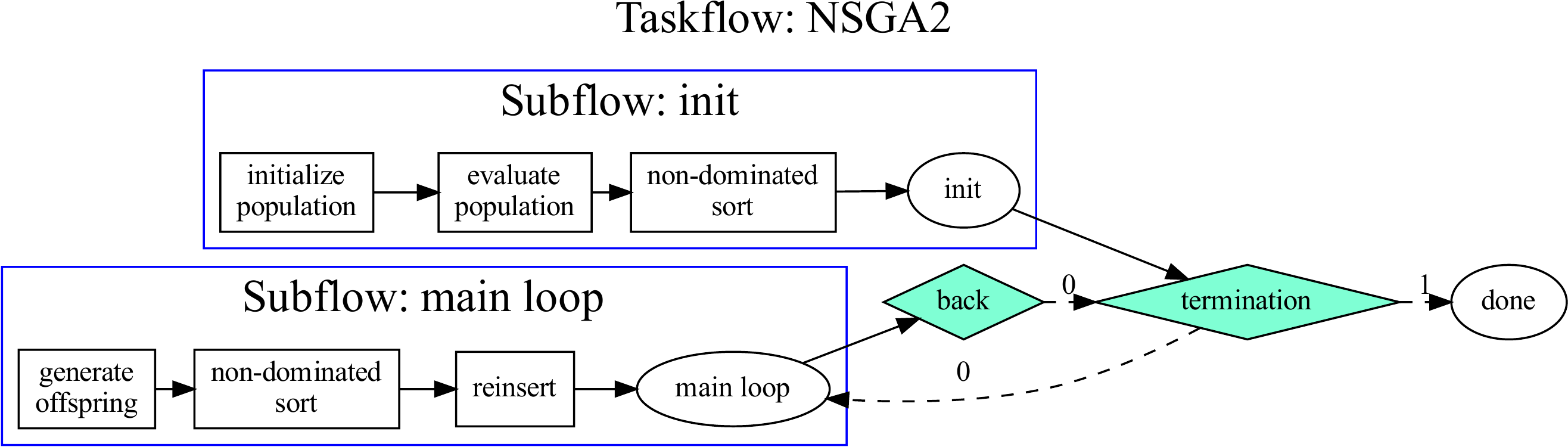}
    \caption{Taskflow describing the NSGA2 algorithm in Operon. Each individual task within the subflows (initialization, evaluation, offspring generation) executes in parallel, using a number of logical threads equal to the population size.}\label{fig:operon-nsga}
\end{figure}

Operon is a \cpp{} framework for symbolic regression that employs logical parallelism during evolution, such that every new offspring individual is generated in its own logical thread. An example evolutionary algorithm implemented in Operon as an operator graph is shown in Figure~\ref{fig:operon-nsga}.
%
%The current version of Operon\footnote{\url{https://github.com/heal-research/operon}} is using Taskflow~\cite{Huang2019} as a back-end for parallelism and implements its evolutionary model as a computational graph, as shown in Figure~\ref{fig:operon-nsga}. The back-end automatically manages threads and distributes work to the CPU cores in an optimal manner.

Operon uses a linear encoding where each tree is represented as a postfix sequence of nodes. Each node has typical attributes such as length, depth, arity or opcode.
%From an implementation standpoint, the \texttt{Node} structure is a ``plain old data`` (POD) type with standard memory layout (it occupies a contiguous memory area), which allows very fast copying operations. This guarantees the efficiency of genetic variation operators such as crossover and mutation where subtrees are swapped or replaced.
%
Evaluation efficiency is achieved by employing a batched tree interpreter, which iterates over the tree nodes and executes the corresponding functions on fixed-size batches of data. As the batch size is known at compile-time, these operations are vectorized. The entire tree evaluation infrastructure relies on the \emph{Eigen} \cpp{} library~\cite{Guennebaud2010} for efficient, vectorized execution.

\subsubsection{Implementing the $\sum$ symbol}\label{subsec:sum-symbol}

The tree interpreter represents a generic approach to tree evaluation and is agnostic of the actual primitive set used by the algorithm. Each node is mapped to a callable\footnote{\url{https://en.wikipedia.org/wiki/Callable_object}} (stateful function object) which defines the functional transformation.
%The specific functions are called upon evaluation via a dispatch table mechanism that maps unique tree node identifiers (hashes) to \emph{callables}\footnote{\url{https://en.wikipedia.org/wiki/Callable_object}} (function objects).
The callables themselves are required to satisfy a certain function signature and to operate in both scalar and dual number domains.
%This is achieved with the help of C\verb!++! templates, such that when a callable is added to the map, the dispatch table conveniently instantiates function wrappers of the callable in all supported number domains.

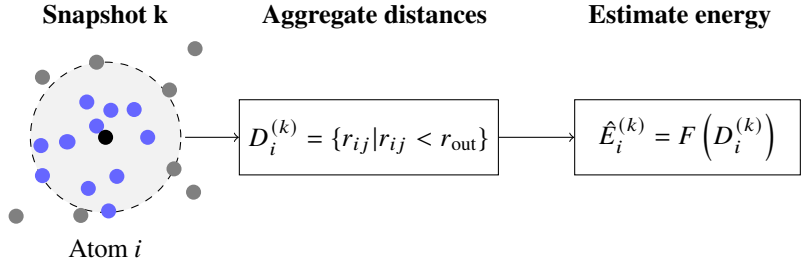
\begin{figure}
    \normalsize
    \input{./sr_potential}
    \caption{Prediction of atom energies using SR. The total energy is then $\displaystyle \hat{E}^{(k)} = \sum_i \hat{E}_i^{(k)}$.}\label{fig:sr-pes}
\end{figure}

This mechanism facilitates the extension of the default primitive set with any kind of ad hoc functionality -- the $\sum$ (summation) symbol in this particular application. Figure~\ref{fig:sr-pes} shows the general workflow for processing a set of atomic positions into pairwise distances and using them to estimate the potential energy. The function $F$ represents a symbolic functional form which includes $\sum$ symbols over pairwise atomic distances.
Since the $\sum$ symbol is essentially a reduction operator\footnote{\url{https://en.wikipedia.org/wiki/Reduction_operator}}, the actual number of dimensions of the input data is three: $N$ snapshots, $M$ atoms, $L$ pairwise distances (where $L$ dynamically depends on the cut-off radius $r_\text{out}$).

Like many other evolutionary frameworks, Operon relies on a dataset object which holds tabular data in two dimensions: $X$ features $\times$ $Y$ observations. Therefore, it is not straightforward to accommodate an additional data dimension without significant redesign work. However, it is relatively easy to incorporate an extra, inner dataset into the function object associated to the $\sum$ symbol, which will contain the values in the third dimension (interatomic distances).
%\footnote{This is possible because function objects are stateful and can store other data.}.

For this mechanism to work, a convention is necessary: the \emph{outer} dataset will contain the target energy values as well as an input variable $r$ whose value is always 1 (this value was chosen arbitrarily as a non-problematic constant which does not cause discontinuities). The variable $r$ simply acts a placeholder for the pairwise atomic distance values. The \emph{inner} dataset will contain the actual pairwise distances under the same input name $r$. The distances are computed when the atomic coordinate values are loaded into the callable. A nested tree interpreter is then used to evaluate the current $\sum$-subtree using the inner dataset as input. Similar to Hernandez et al.~\cite{Hernandez2019}, the $\sum$ symbol also applies a smoothing function on its output (see Equation (7) in~\cite{Hernandez2019}) with the inner and outer cutoff radii equal to $3 \text{\r{A}}$ and $5 \text{\r{A}}$, respectively.

Under this set of rules, a leaf node corresponding to the input variable $r$ will evaluate to 1 when not under a $\sum$ symbol, and to the set of pairwise distance values corresponding to the current atomic configuration otherwise. Additionally, to disallow nesting of $\sum$ symbols, the behavior is dynamically switched depending on the surrounding tree context: if a $\sum$ symbol finds itself under another $\sum$ symbol, then it simply acts as the identity function $f(x) = x$.
This convention does not impact the evolutionary system's ability to discover interatomic potential functional forms.

\subsection{Empirical validation}\label{subsec:empirical}

We demonstrate the capabilities of the proposed NSGA2-based multi-objective approach using the \abinitio{} data used by Hernandez et al.~\cite{Hernandez2019}. This data consists of $150$ snapshots of 32-atom DFT molecular dynamics simulations of copper (Cu): 50 snapshots at 300 K (NVT), 50 snapshots at 1400 K (NVT) and 50 snapshots at 1400 K (NPT at 100 kPa). Although the data also contains components of forces and virial stress tensors, only the energy was used as a modeling target in this experiment. The data consisting of 150 configurations is shuffled and split equally into training and test partitions.

\subsubsection{Experimental setup}\label{subsubsec:experimental-setup}

%All tests were performed on a workstation containing an AMD Ryzen\texttrademark{} 5950X CPU with 8Mb of L2 cache and 64Mb of L3 cache, and 64Gb of DDR4 3600Mhz-CL16 memory.
%
The experiment used a fixed set of parameters shown in Table~\ref{tbl:fixed-parameters}.
The primitive set was varied and consisted of different symbol combinations, as shown in Table~\ref{tbl:results}: with and without the power function, and alternating between $\div$ and $\text{aq}$, where $\text{aq}(x) = \sqrt{x^2 + 1}$.

\begin{table}
    \centering
    \newcommand{\tabitem}{~~\llap{\textbullet}~~}
    \begin{tabular}{ll}
        population size & 10,000 individuals\\
        tree initialization & balanced tree creator (BTC)~\cite{Burlacu2020}\\
        max tree length & 20\\
        max tree depth & 10\\
        crossover probability & 100\%\\
        crossover operator & subtree crossover\\
        mutation probability & 25\%\\
        mutation operator & uniformly chosen from:\\
                          & \tabitem subtree removal/insertion/replacement\\
                          & \tabitem change function symbol\\
                          & \tabitem change variable name\\
                          & \tabitem additive one point leaf mutation ($v = v + \mathcal{N}(0,1)$)\\
                          & \tabitem discrete point leaf mutation ($v \gets$ math constant: $\pi, \E$, ...)\\
        selection operator & crowded tournament selection, group size = 17\\
        objectives & Pearson R$^2$ and model length\\
        evaluation budget & $10^8$ fitness evaluations\\
    \end{tabular}
    \caption{NSGA2 parameters}\label{tbl:fixed-parameters}
\end{table}

Two input variables are used: $r$ as a placeholder for atomic distances and $q = \frac{1}{r}$ as a placeholder for the inverse of $r$, given that some empirical potentials like Lennard-Jones explicitly use the inverse in their formula.
Each experimental configuration was repeated 50 times and the median values were reported (with the exception of runtime, which was averaged). Errors are reported as median $\pm$ standard deviation. Model length was computed as the length of the simplified representation returned by \emph{Sympy}, using the infix textual representation of the best individual as input.

%\begin{equation}
%    \text{Fitness} = \frac{1}{N} \sum_{k=1}^{N} \left| E^{(k)} - \hat{E}^{(k)} \right|
%\end{equation}

\subsubsection{Results}\label{subsubsec:results}

Results aggregated over 50 runs for each configuration are shown in Table~\ref{tbl:results}, alongside p-value matrices computed using the Kruskal test. Significance is encoded in Tables~\ref{tbl:train-significance} and~\ref{tbl:test-significance} using font weight and color: values lower than $\alpha=0.01$ are shown in bold black font, values lower than $\alpha=0.05$ are shown in black font, while all the other values are shown in gray. The direction of the relationship is determined using a comparison of median values and shown as $\uparrow$ (worse/higher error) or $\downarrow$ (better/lower error) symbols prefixed to the values.

The overall best models from all runs and all configurations are shown in Table~\ref{tbl:best-models}. These models have been selected based on both test accuracy and simplicity of their functional form. Two other models with better test score have been discarded due to very complex structure or very large coefficient values. Table~\ref{tbl:best-models} illustrates this fact by displaying the absolute rank of each model (based purely on test accuracy and disregarding other criteria).

Interestingly, the arithmetic-only configurations A, B, C generated 4 out of 5 of the selected best models. Although configuration A produced significantly worse ($p < 0.01$) training accuracies than all other configurations, it did not produce worse models in terms of generalization, where it is only worse than E. Nevertheless, the explicit inclusion of $1/r$ as an input seems to help the search.

It is also worth noting that configurations using the analytic quotient instead of (unprotected) division generally perform better on the training data ($p < 0.05$), but do not perform better on the test data. For example, configuration K is better than A, B, C, H, I in terms of training accuracy, but is not better than any of them in terms of test (on the contrary, it is worse than E at $p < 0.05$). From this we can surmise that in this particular test setting and for this particular data, AQ does not offer an advantage compared to normal division.

Overall, judging from median error values and statistical significance p-values, there is no clear winner among the tested configurations. However, a pattern emerges when observing the functional forms of the best models, mostly originating from configurations B and C. After simplification using \emph{Sympy}, the models become highly similar with the same mathematical structure consisting of a sum of three factors in the numerator (each including the inverse of $r$) and another sum in the denominator (also including the inverse of $r$). Although these models are remarkably simple, further testing is required to validate their properties and behavior.

In terms of runtime, the proposed approach is efficient, with the longest run taking on average 290 seconds to evolve a population of 10,000 individuals for 1000 generations on a single multicore computer. In comparison, Hernandez et al.~\cite{Hernandez2019} report 360 CPU-hours expended on finding accurate GP models.

\begin{table}
    \setlength{\tabcolsep}{4.7pt}
    \include{./aggregated_results}

    \caption{Operon NSGA2 Results}\label{tbl:results}
\end{table}

\begin{table}
    \setlength{\tabcolsep}{3.4pt}
    \scriptsize
    \include{./sig_train}
\caption{Training error p-value matrix using the Kruskal statistical test. Significance shown by bold black font ($p<0.01$), black font ($p < 0.05$) or gray (no significance). Relationship direction given by comparison of medians: $\uparrow$ (worse/higher error), $\downarrow$ (better/lower error).}\label{tbl:train-significance}
\end{table}

\begin{table}
    \setlength{\tabcolsep}{3.4pt}
    \scriptsize
    \include{./sig_test}
    \caption{Test error p-value matrix using the Kruskal statistical test. Significance shown by bold black font ($p<0.01$), black font ($p < 0.05$) or gray (no significance). Relationship direction given by comparison of medians: $\uparrow$ (worse/higher error), $\downarrow$ (better/lower error).}\label{tbl:test-significance}
\end{table}

\begin{table}
    \include{./best_models}

    \caption{Overall best models, where ID identifies the configuration in Table~\ref{tbl:results}.}\label{tbl:best-models}
\end{table}

\section{Conclusion}

This work surveyed the main applications of SR in Materials Science, namely for the discovery of simple and efficient models of interatomic potentials. Both previous results, as well as results obtained by our own proposed approach and described in this paper, suggest that SR is capable of finding accurate models that can further the capabilities of particle simulations.

Similar to POET~\cite{Hernandez2019}, our approach does not restrict the search space in any way (with the exception of tree length and depth limits) and is therefore capable of finding models that do not resemble previously known, empirical potential functions. At the same time, should a directed search be required, the framework is trivial to extend with this feature.

Empirical testing shows that relatively simple primitive sets are powerful enough to discover accurate potential functions with good extrapolation behavior. On this data, no advantage was found in using the analytical quotient over standard division. More experiments will be required to establish the benefits of larger primitive sets, for example ones that include logarithmic, exponential or trigonometric functions.

Several other aspects like a more comprehensive search in the space of hyper-parameters or an exploration of the effects of local search also need to be fully investigated in the future. Compared to other works described in our survey, our approach did not diverge from the ``vanilla'' version of GP, using a classical multi-objective approach (NSGA2) together with a domain specific primitive set. It will be also worthwhile to explore various ways to scale up the search using multiple populations and more sophisticated evolutionary models.

Future development directions include expanding the capabilities of the framework to include three- or many-body interactions, to consider model derivatives in order model atomic forces as well, and overall to improve its ability to incorporate and respect the fundamental laws of this kind of physical systems.

\newpage
\section{Appendix}\label{sec:appendix}

\subsection*{Empirical potentials}\label{subsec:empirical-potentials}

For a comprehensive overview of empirical potentials we recommend the work of Ara\'{u}jo and Ballester~\cite{Araujo2021}. Below we give a casual overview of the most important empirical potentials mentioned in this contribution.

\subsubsection*{Morse potential}
This is an empirical potential used to model diatomic molecules.

\begin{equation}
    V_\text{M}(r) = D\Big( 1 - \exp \big( -a(r - r_0) \big) ^2\Big)\label{eq:morse-potential}
\end{equation}
where $D$ is the dissociation energy, $r$ is the distance between atoms, $a$ is a set of parameters and $r_0$ is the equilibrium bond distance.

\subsubsection*{Lennard-Jones potential}

The Lennard-Jones potential models soft repulsive and attractive interactions and can describe electronically neutral atoms or molecules. Interacting particles repel each other at very close distance, attract each other at moderate distance, and do not interact at infinite distance.

\begin{equation}
    V_\text{LJ}(r) = 4 \varepsilon \bigg[ \Big( \frac{\sigma}{r} \Big)^{12} - \Big( \frac{\sigma}{r} \Big)^6 \bigg]\label{eq:lennard-jones-potential}
\end{equation}
where $r$ is the distance between atoms, $\varepsilon$ is the dispersion energy and $\sigma$ is the distance at which the particle-particle potential energy $V$ is zero.

\subsubsection*{Lippincott potential}

Lippincott~\cite{Steele1962} potential involves an exponential of interatomic distances
\begin{equation}
    V_\text{LIP}(r) = D\Bigg(1 - \exp \Big(\frac{-n(r - r_0)^2}{2r} \Big) \Bigg)\Big(1 + aF(r) \Big)\label{eq:lippincott-potential}
\end{equation}
where $D$ is the dissociation energy, $r$ is the distance between atoms, $r_0$ is the equilibrium bond distance and $a$ and $n$ are parameters. $F(r)$ is a function of internuclear distance such that $F(r) = 0$ when $r=\infty$ and $F(r) = \infty$ when $r=0$.
%\begin{equation}
%    V_L(r) = D(1 - e^{-x})\Big(1 - ab\sqrt{x} \exp\big( -b\sqrt{xr_0/r} \big) \Big)\label{eq:lippincott-potential}
%\end{equation}

\subsubsection*{Stillinger-Weber potential}

The Stillinger-Weber potential~\cite{Stillinger1985} models two- and three-body interactions by taking into account not only the distances between atoms but also the bond angles:

\begin{equation}
    V_\text{SW}(r) = \sum_{\langle i, j \rangle} \phi_2(r_{ij}) + \sum_{\langle i, j, k \rangle} \phi_3(r_{ij}, r_{ik}, \theta_{ijk})\label{eq:stillinger-weber-potential}
\end{equation}
where
\begin{align}
    \phi_2(r_{ij}) &= A \varepsilon \left[ B\left( \frac{\sigma}{r_{ij}} \right)^p - \left( \frac{\sigma}{r_{ij}} \right)^q \right] \exp \left( \frac{\sigma}{r_{ij} - a\sigma} \right)\text{ and}\\
    \phi_3(r_{ij}, r_{ik}, \theta_{ijk}) &= \lambda \varepsilon \left[ \cos \theta_{ijk} - \cos \theta_0 \right]^2 \times \exp \left( \frac{\gamma \sigma}{r_{ij} - a\sigma} \right) \exp \left( \frac{\gamma \sigma}{r_{ik} - a \sigma} \right)
\end{align}

\subsubsection*{Sutton-Chen potential}

The Sutton-Chen potential~\cite{Sutton1990} has been used in molecular dynamics and Monte Carlo simulations of metallic systems. It offers a reasonable description of various bulk properties, with an approximate many-body representation of the delocalized metallic bonding:
\begin{equation}
    V_\text{SC} = \sum_{\langle i,j \rangle} U(r_{ij}) - \sum_i u\sqrt{\rho_i}
\end{equation}
Here, the first term represents the repulsion between atomic cores and the second term models the bonding energy due to the electrons. Both terms are further defined in terms of reciprocal power so that the complete expression is:
\begin{equation}
    V_\text{SC} = \epsilon \left[ \sum_{\langle i,j \rangle} \left( \frac{a}{r_{ij}} \right)^n - C \sum_i \sqrt{\sum_j \left( \frac{a}{r_{ij}} \right)^m} \right]
\end{equation}
where $C$ is a dimensionless parameter, $\epsilon$ is a parameter with dimensions of energy, $a$ is the lattice constant, $m, n$ are positive integers with $n > m$ and $r_{ij}$ is the distance between the $i$th and $j$th atoms.

\bibliography{references.bib}
\bibliographystyle{plain}

\end{document}

%% file: poet-example.tex
\begin{tikzpicture}[every node/.style={circle,draw},level 1/.style={sibling distance=2cm, level distance=1cm}]
\node {$+$}
    child [xshift=-1cm] { node {$\times$}
        child { node {$\sum$}
            child { node {$\times$}
                child { node { $\land$ }
                    child { node [rectangle] { $r$ } }
                    child { node { $-$ }
                        child { node [rectangle] { $3.98$ } }
                        child { node { $\times$ }
                            child { node [rectangle] { $3.93$ } }
                            child { node [rectangle] { $r$ } }
                        }
                    }
                }
                child { node [rectangle] { $f(r)$ } }
            }
        }
        child { node [rectangle] { $7.51$ } }
    }
    child [xshift=1cm] { node {$\div$}
        child { node { $-$ }
            child { node [rectangle] { $28.0$ } }
            child { node { $\times$ }
                child { node [rectangle] { $0.03$ } }
                child { node { $\sum$ }
                    child { node { $\times$ }
                        child { node { $\land$ }
                            child { node { $r$ } }
                            child { node { $-$ }
                                child { node [rectangle] { $11.7$ } }
                                child { node { $\times$ }
                                    child { node [rectangle] { $2.93$ } }
                                    child { node [rectangle] { $r$ } }
                                }
                            }
                        }
                        child { node [rectangle] { $f(r)$ } }
                    }
                }
            }
        }
        child { node { $\sum$ }
            child { node [rectangle] { $f(r)$ } }
        }
    };
\end{tikzpicture}

%% file: sr_potential.tex
\begin{tikzpicture}
    % Random seed for RNG
    \pgfmathsetseed{\number\pdfrandomseed}

    % Define circle parameters
    \newcommand{\cX}{0}
    \newcommand{\cY}{0}
    \newcommand{\cR}{1}
    \newcommand{\cD}{2.1}

    %\draw (0,0) -- (4,0) -- (4,4) -- (0,4) -- cycle;

    \draw[dashed,fill=gray!10!white, minimum width=\cD cm] (\cX,\cY) circle (\cR) node (atom) {};
    \foreach \x in {1,...,20}
    {
      % Find random numbers
      \pgfmathrandominteger{\a}{-120}{120}
      \pgfmathrandominteger{\b}{-120}{120}

      % Scale numbers nicely
      \pgfmathparse{0.01*\a}\let\a\pgfmathresult
      \pgfmathparse{0.01*\b}\let\b\pgfmathresult

      % Check if numbers are inside circle
      \pgfmathparse{ifthenelse((\a-\cX)^2 + (\b-\cY)^2 < \cR^2,%
          "blue!60!white",
          "gray")}
        \fill[\pgfmathresult] (\a,\b) circle (0.1);
    };
    \fill[black] (0, 0) circle (0.1);
    \node[below=of atom, yshift=-1mm] {Atom $i$ };
    \node[above=of atom, yshift=2mm] (snapshot) {\textbf{Snapshot} $\mathbf{k}$};
    \node[draw, right of=atom, xshift=2.5cm, minimum height=1cm] (dist) { $D_i^{(k)} = \{ r_{ij} | r_{ij} < r_\text{out} \}$ };
    \node[right=of snapshot, xshift=0cm] (agg) { \textbf{Aggregate distances} };
    \node[right=of dist,draw, minimum height=1cm, minimum width=3cm] (model) { $\hat{E}_i^{(k)} = F\left(D_i^{(k)}\right)$ };
    \node[right=of agg, xshift=0.1cm, align=left] { \textbf{Estimate energy} };

    \draw[->] (atom) -- (dist);
    \draw[->,xshift=1cm] (dist) -- (model);
\end{tikzpicture}

%% file: aggregated_results.tex
\begin{tabular}{lllcccc}
    \textbf{ID} & \textbf{Primitive set} & \textbf{Inputs} & \textbf{MAE$_\text{train}$} & \textbf{MAE$_\text{test}$} & \textbf{Length} & \textbf{Runtime (s)}\\
    \toprule                                                                                                     % Experiment id:
    A & $\sum,+,-,\times,\div$ & $r$ & $0.568 \pm 0.045$ & $0.602 \pm 0.059$ & $32.0$ & $118.52$\\                   % f9d1dc1d3530104fc270
    B & $\sum,+,-,\times,\div$ & $q$ & $0.518 \pm 0.036$ & $0.599 \pm 0.069$ & $44.0$ & $142.01$\\                   % 4cb825ade74380b07958
    C & $\sum,+,-,\times,\div$ & $r, q$ & $0.512 \pm 0.043$ & $0.595 \pm 0.091$ & $42.0$ & $143.69$\\                % f53b3d76fcdd07ed9be7
    \midrule
    D & $\sum,+,-,\times,\text{aq}$ & $r$ & $0.498 \pm 0.047$ & $0.583 \pm 0.060$ & $56.0$ & $165.49$\\              % e1b097bc41e90ad5a62e
    E & $\sum,+,-,\times,\text{aq}$ & $q$ & $0.500 \pm 0.066$ & $0.574 \pm 0.068$ & $56.5$ & $162.51$\\              % 14538a92c23403491b5c
    F & $\sum,+,-,\times,\text{aq}$ & $r, q$ & $0.493 \pm 0.046$ & $0.593 \pm 0.060$ & $60.0$ & $169.64$\\           % 9a5ee69b6a341db00a6a
    \midrule
    G & $\sum,+,-,\times,\div,\text{pow}$ & $r$ & $0.501 \pm 0.042$ & $0.620 \pm 0.039$ & $39.0$  & $286.95$\\       % 254c0b27468dc7a79760
    H & $\sum,+,-,\times,\div,\text{pow}$ & $q$ & $0.516 \pm 0.048$ & $0.604 \pm 0.065$ & $46.5$ & $241.25$\\        % a376ffd790815c616500
    I & $\sum,+,-,\times,\div,\text{pow}$ & $r, q$ & $0.514 \pm 0.051$ & $0.596 \pm 0.057$ & $47.0$ & $290.53$\\     % a2615fbe632a5844452b
    \midrule
    J & $\sum,+,-,\times,\text{aq},\text{pow}$ & $r$ & $0.507 \pm 0.052$ & $0.608 \pm 0.059$ & $47.0$ & $269.26$\\   % 688727cad861a61a3318
    K & $\sum,+,-,\times,\text{aq},\text{pow}$ & $q$ & $0.489 \pm 0.053$ & $0.623 \pm 0.085$ & $57.0$ & $244.44$\\   % 52b9fb6ed7339aaa95e5
    L & $\sum,+,-,\times,\text{aq},\text{pow}$ & $r, q$ & $0.497 \pm 0.053$ & $0.594\pm 0.068$ & $57.0$ & $281.86$\\ % c2e0c27465978d12a921
    \midrule
\end{tabular}

%% file: sig_train.tex
\begin{tabular}{ccccccccccccc}
 & A & B & C & D & E & F & G & H & I & J & K & L \\
    \addlinespace
A & \color{transparent}  & \color{black} \bfseries $\uparrow$4e-07 & \color{black} \bfseries $\uparrow$8e-07 & \color{black} \bfseries $\uparrow$1e-09 & \color{black} \bfseries $\uparrow$2e-07 & \color{black} \bfseries $\uparrow$1e-09 & \color{black} \bfseries $\uparrow$1e-08 & \color{black} \bfseries $\uparrow$3e-06 & \color{black} \bfseries $\uparrow$8e-06 & \color{black} \bfseries $\uparrow$3e-07 & \color{black} \bfseries $\uparrow$9e-10 & \color{black} \bfseries $\uparrow$9e-09 \\
    \addlinespace
B & \color{black} \bfseries$\downarrow$4e-07 & \color{transparent}  & \color{gray} $\uparrow$6e-01 & \color{black} $\uparrow$2e-02 & \color{gray} $\uparrow$1e-01 & \color{black} $\uparrow$2e-02 & \color{gray} $\uparrow$1e-01 & \color{gray} $\uparrow$1e+00 & \color{gray} $\uparrow$8e-01 & \color{gray} $\uparrow$3e-01 & \color{black} \bfseries $\uparrow$6e-03 & \color{black} $\uparrow$5e-02 \\
    \addlinespace
C & \color{black} \bfseries$\downarrow$8e-07 & \color{gray}$\downarrow$6e-01 & \color{transparent}  & \color{gray} $\uparrow$7e-02 & \color{gray} $\uparrow$2e-01 & \color{black} $\uparrow$5e-02 & \color{gray} $\uparrow$3e-01 & \color{gray}$\downarrow$7e-01 & \color{gray}$\downarrow$9e-01 & \color{gray} $\uparrow$5e-01 & \color{black} $\uparrow$2e-02 & \color{gray} $\uparrow$1e-01 \\
    \addlinespace
D & \color{black} \bfseries$\downarrow$1e-09 & \color{black}$\downarrow$2e-02 & \color{gray}$\downarrow$7e-02 & \color{transparent}  & \color{gray}$\downarrow$4e-01 & \color{gray} $\uparrow$9e-01 & \color{gray}$\downarrow$3e-01 & \color{black}$\downarrow$4e-02 & \color{gray}$\downarrow$6e-02 & \color{gray}$\downarrow$3e-01 & \color{gray} $\uparrow$7e-01 & \color{gray} $\uparrow$8e-01 \\
    \addlinespace
E & \color{black} \bfseries$\downarrow$2e-07 & \color{gray}$\downarrow$1e-01 & \color{gray}$\downarrow$2e-01 & \color{gray} $\uparrow$4e-01 & \color{transparent}  & \color{gray} $\uparrow$5e-01 & \color{gray}$\downarrow$8e-01 & \color{gray}$\downarrow$1e-01 & \color{gray}$\downarrow$3e-01 & \color{gray}$\downarrow$9e-01 & \color{gray} $\uparrow$2e-01 & \color{gray} $\uparrow$7e-01 \\
    \addlinespace
F & \color{black} \bfseries$\downarrow$1e-09 & \color{black}$\downarrow$2e-02 & \color{black}$\downarrow$5e-02 & \color{gray}$\downarrow$9e-01 & \color{gray}$\downarrow$5e-01 & \color{transparent}  & \color{gray}$\downarrow$4e-01 & \color{black}$\downarrow$2e-02 & \color{gray}$\downarrow$6e-02 & \color{gray}$\downarrow$3e-01 & \color{gray} $\uparrow$6e-01 & \color{gray}$\downarrow$8e-01 \\
    \addlinespace
G & \color{black} \bfseries$\downarrow$1e-08 & \color{gray}$\downarrow$1e-01 & \color{gray}$\downarrow$3e-01 & \color{gray} $\uparrow$3e-01 & \color{gray} $\uparrow$8e-01 & \color{gray} $\uparrow$4e-01 & \color{transparent}  & \color{gray}$\downarrow$1e-01 & \color{gray}$\downarrow$3e-01 & \color{gray}$\downarrow$8e-01 & \color{gray} $\uparrow$2e-01 & \color{gray} $\uparrow$6e-01 \\
    \addlinespace
H & \color{black} \bfseries$\downarrow$3e-06 & \color{gray}$\downarrow$1e+00 & \color{gray} $\uparrow$7e-01 & \color{black} $\uparrow$4e-02 & \color{gray} $\uparrow$1e-01 & \color{black} $\uparrow$2e-02 & \color{gray} $\uparrow$1e-01 & \color{transparent}  & \color{gray} $\uparrow$8e-01 & \color{gray} $\uparrow$3e-01 & \color{black} \bfseries $\uparrow$7e-03 & \color{black} $\uparrow$5e-02 \\
    \addlinespace
I & \color{black} \bfseries$\downarrow$8e-06 & \color{gray}$\downarrow$8e-01 & \color{gray} $\uparrow$9e-01 & \color{gray} $\uparrow$6e-02 & \color{gray} $\uparrow$3e-01 & \color{gray} $\uparrow$6e-02 & \color{gray} $\uparrow$3e-01 & \color{gray}$\downarrow$8e-01 & \color{transparent}  & \color{gray} $\uparrow$4e-01 & \color{black} $\uparrow$3e-02 & \color{gray} $\uparrow$1e-01 \\
    \addlinespace
J & \color{black} \bfseries$\downarrow$3e-07 & \color{gray}$\downarrow$3e-01 & \color{gray}$\downarrow$5e-01 & \color{gray} $\uparrow$3e-01 & \color{gray} $\uparrow$9e-01 & \color{gray} $\uparrow$3e-01 & \color{gray} $\uparrow$8e-01 & \color{gray}$\downarrow$3e-01 & \color{gray}$\downarrow$4e-01 & \color{transparent}  & \color{gray} $\uparrow$1e-01 & \color{gray} $\uparrow$5e-01 \\
    \addlinespace
K & \color{black} \bfseries$\downarrow$9e-10 & \color{black} \bfseries$\downarrow$6e-03 & \color{black}$\downarrow$2e-02 & \color{gray}$\downarrow$7e-01 & \color{gray}$\downarrow$2e-01 & \color{gray}$\downarrow$6e-01 & \color{gray}$\downarrow$2e-01 & \color{black} \bfseries$\downarrow$7e-03 & \color{black}$\downarrow$3e-02 & \color{gray}$\downarrow$1e-01 & \color{transparent}  & \color{gray}$\downarrow$4e-01 \\
    \addlinespace
L & \color{black} \bfseries$\downarrow$9e-09 & \color{black}$\downarrow$5e-02 & \color{gray}$\downarrow$1e-01 & \color{gray}$\downarrow$8e-01 & \color{gray}$\downarrow$7e-01 & \color{gray} $\uparrow$8e-01 & \color{gray}$\downarrow$6e-01 & \color{black}$\downarrow$5e-02 & \color{gray}$\downarrow$1e-01 & \color{gray}$\downarrow$5e-01 & \color{gray} $\uparrow$4e-01 & \color{transparent}  \\
\end{tabular}

%% file: sig_test.tex
\begin{tabular}{ccccccccccccc}
 & A & B & C & D & E & F & G & H & I & J & K & L \\
\addlinespace
A & \color{transparent}  & \color{gray} $\uparrow$3e-01 & \color{gray} $\uparrow$4e-01 & \color{gray} $\uparrow$2e-01 & \color{black} \bfseries $\uparrow$7e-03 & \color{gray} $\uparrow$8e-02 & \color{gray}$\downarrow$9e-01 & \color{gray}$\downarrow$4e-01 & \color{gray} $\uparrow$1e-01 & \color{gray}$\downarrow$5e-01 & \color{gray}$\downarrow$9e-01 & \color{gray} $\uparrow$7e-02 \\
\addlinespace
B & \color{gray}$\downarrow$3e-01 & \color{transparent}  & \color{gray} $\uparrow$9e-01 & \color{gray} $\uparrow$1e+00 & \color{gray} $\uparrow$2e-01 & \color{gray} $\uparrow$6e-01 & \color{gray}$\downarrow$2e-01 & \color{gray}$\downarrow$7e-01 & \color{gray} $\uparrow$6e-01 & \color{gray}$\downarrow$8e-02 & \color{gray}$\downarrow$2e-01 & \color{gray} $\uparrow$4e-01 \\
\addlinespace
C & \color{gray}$\downarrow$4e-01 & \color{gray}$\downarrow$9e-01 & \color{transparent}  & \color{gray} $\uparrow$1e+00 & \color{gray} $\uparrow$2e-01 & \color{gray} $\uparrow$5e-01 & \color{gray}$\downarrow$4e-01 & \color{gray}$\downarrow$8e-01 & \color{gray}$\downarrow$8e-01 & \color{gray}$\downarrow$1e-01 & \color{gray}$\downarrow$3e-01 & \color{gray} $\uparrow$6e-01 \\
\addlinespace
D & \color{gray}$\downarrow$2e-01 & \color{gray}$\downarrow$1e+00 & \color{gray}$\downarrow$1e+00 & \color{transparent}  & \color{gray} $\uparrow$1e-01 & \color{gray}$\downarrow$6e-01 & \color{gray}$\downarrow$3e-01 & \color{gray}$\downarrow$8e-01 & \color{gray}$\downarrow$9e-01 & \color{black}$\downarrow$4e-02 & \color{gray}$\downarrow$3e-01 & \color{gray}$\downarrow$6e-01 \\
\addlinespace
E & \color{black} \bfseries$\downarrow$7e-03 & \color{gray}$\downarrow$2e-01 & \color{gray}$\downarrow$2e-01 & \color{gray}$\downarrow$1e-01 & \color{transparent}  & \color{gray}$\downarrow$4e-01 & \color{black} \bfseries$\downarrow$3e-03 & \color{gray}$\downarrow$8e-02 & \color{gray}$\downarrow$3e-01 & \color{black} \bfseries$\downarrow$1e-03 & \color{black}$\downarrow$1e-02 & \color{gray}$\downarrow$4e-01 \\
\addlinespace
F & \color{gray}$\downarrow$8e-02 & \color{gray}$\downarrow$6e-01 & \color{gray}$\downarrow$5e-01 & \color{gray} $\uparrow$6e-01 & \color{gray} $\uparrow$4e-01 & \color{transparent}  & \color{black}$\downarrow$5e-02 & \color{gray}$\downarrow$4e-01 & \color{gray}$\downarrow$7e-01 & \color{black}$\downarrow$1e-02 & \color{gray}$\downarrow$8e-02 & \color{gray}$\downarrow$9e-01 \\
\addlinespace
G & \color{gray} $\uparrow$9e-01 & \color{gray} $\uparrow$2e-01 & \color{gray} $\uparrow$4e-01 & \color{gray} $\uparrow$3e-01 & \color{black} \bfseries $\uparrow$3e-03 & \color{black} $\uparrow$5e-02 & \color{transparent}  & \color{gray} $\uparrow$3e-01 & \color{black} $\uparrow$3e-02 & \color{gray} $\uparrow$5e-01 & \color{gray}$\downarrow$7e-01 & \color{black} $\uparrow$2e-02 \\
\addlinespace
H & \color{gray} $\uparrow$4e-01 & \color{gray} $\uparrow$7e-01 & \color{gray} $\uparrow$8e-01 & \color{gray} $\uparrow$8e-01 & \color{gray} $\uparrow$8e-02 & \color{gray} $\uparrow$4e-01 & \color{gray}$\downarrow$3e-01 & \color{transparent}  & \color{gray} $\uparrow$4e-01 & \color{gray}$\downarrow$1e-01 & \color{gray}$\downarrow$3e-01 & \color{gray} $\uparrow$3e-01 \\
\addlinespace
I & \color{gray}$\downarrow$1e-01 & \color{gray}$\downarrow$6e-01 & \color{gray} $\uparrow$8e-01 & \color{gray} $\uparrow$9e-01 & \color{gray} $\uparrow$3e-01 & \color{gray} $\uparrow$7e-01 & \color{black}$\downarrow$3e-02 & \color{gray}$\downarrow$4e-01 & \color{transparent}  & \color{black}$\downarrow$2e-02 & \color{gray}$\downarrow$1e-01 & \color{gray} $\uparrow$7e-01 \\
\addlinespace
J & \color{gray} $\uparrow$5e-01 & \color{gray} $\uparrow$8e-02 & \color{gray} $\uparrow$1e-01 & \color{black} $\uparrow$4e-02 & \color{black} \bfseries $\uparrow$1e-03 & \color{black} $\uparrow$1e-02 & \color{gray}$\downarrow$5e-01 & \color{gray} $\uparrow$1e-01 & \color{black} $\uparrow$2e-02 & \color{transparent}  & \color{gray}$\downarrow$7e-01 & \color{black} \bfseries $\uparrow$7e-03 \\
\addlinespace
K & \color{gray} $\uparrow$9e-01 & \color{gray} $\uparrow$2e-01 & \color{gray} $\uparrow$3e-01 & \color{gray} $\uparrow$3e-01 & \color{black} $\uparrow$1e-02 & \color{gray} $\uparrow$8e-02 & \color{gray} $\uparrow$7e-01 & \color{gray} $\uparrow$3e-01 & \color{gray} $\uparrow$1e-01 & \color{gray} $\uparrow$7e-01 & \color{transparent}  & \color{gray} $\uparrow$7e-02 \\
\addlinespace
L & \color{gray}$\downarrow$7e-02 & \color{gray}$\downarrow$4e-01 & \color{gray}$\downarrow$6e-01 & \color{gray} $\uparrow$6e-01 & \color{gray} $\uparrow$4e-01 & \color{gray} $\uparrow$9e-01 & \color{black}$\downarrow$2e-02 & \color{gray}$\downarrow$3e-01 & \color{gray}$\downarrow$7e-01 & \color{black} \bfseries$\downarrow$7e-03 & \color{gray}$\downarrow$7e-02 & \color{transparent}  \\
\end{tabular}

%% file: best_models.tex
\begin{tabular}{p{0.1\textwidth}p{0.88\textwidth}}
    \textbf{ID} & \textbf{Model}\\
    \toprule
    \textbf{C}  & $\text{MAE}_\text{train} = 0.579$, $\text{MAE}_\text{test} = 0.448$, Absolute rank: 1\\
                & $\displaystyle -110.531 - \frac{2929.411 \sum{\left(\left(-0.974 + \frac{2.68}{r}\right) \left(0.727 - \frac{2.888}{r}\right) \left(0.727 - \frac{1.747}{r}\right) \right)}}{\sum{\left(- \frac{0.972}{r \left(0.899 r - 1.815\right)} \right)}} $\\ % 4375046024
    \addlinespace
    \textbf{E}  & $\text{MAE}_\text{train} = 0.612$, $\text{MAE}_\text{test} = 0.454$, Absolute rank: 2\\
                & $\displaystyle \frac{3.037 \left(- \sum{\left(\frac{0.211}{r^{2}} \right)} - 2.396\right)}{\sqrt{\sum^{2}{\left(\left(-2.409 + \frac{6.254}{r}\right) \left(1.209 - \frac{4.99}{r}\right) \left(1.209 - \frac{2.956}{r}\right) \right)} + 1}} - 101.086$\\ % 1609769036
    \addlinespace
    \textbf{C}  & $\text{MAE}_\text{train} = 0.585$, $\text{MAE}_\text{test} = 0.458$, Absolute rank: 3\\
      & $\displaystyle -111.611 + \frac{12327.356 \sum{\left(\left(-0.817 + \frac{2.014}{r}\right) \left(0.318 - \frac{1.255}{r}\right) \left(0.706 - \frac{1.913}{r} \right) \right)}}{\sum{\left(\frac{0.806}{r \left(0.307 r - \frac{1.292}{r}\right)} \right)}}$\\ % 5594889616
    \addlinespace
    \textbf{C} & $\text{MAE}_\text{train} = 0.550$, $\text{MAE}_\text{test} = 0.473$, Absolute rank: 6\\
      & $\displaystyle -108.409 + \frac{14618.749 \sum{\left(\left(0.555 - \frac{1.538}{r}\right) \left(0.707 - \frac{1.667}{r}\right) \left(0.787 - \frac{3.116}{r}\right) \right)}}{\sum{\left(\frac{3.142}{r \left(1.922 - 0.953 r\right)} \right)}}$\\ % 8021701458
    \addlinespace
    \textbf{B} & $\text{MAE}_\text{train} = 0.549$, $\text{MAE}_\text{test} = 0.475$, Absolute rank: 7\\
      & $\displaystyle -109.903 - \frac{82734.094 \sum{\left(\left(-0.361 + \frac{1.414}{r}\right) \left(0.527 - \frac{1.433}{r}\right) \left(0.622 - \frac{1.512}{r}\right) \right)}}{\sum{\left(\frac{0.873}{r \left(-0.339 + \frac{0.686}{r}\right)} \right)}}$\\ % 9807443921
\end{tabular}